\documentclass[12pt]{iopart}

\usepackage{graphicx}% Include figure files
\usepackage{dcolumn}% Align table columns on decimal point
\usepackage{bm}% bold math
\usepackage{cite}
\usepackage{amssymb}
\begin{document}
%\preprint{APS/123-QED}

\title{The impact of kinetic effects on the properties of relativistic electron-positron shocks}

\author{Anne Stockem$^{1}$, Frederico Fi\'{u}za$^{1}$, Ricardo A. Fonseca$^{1,2}$ and Luis O. Silva$^{1}$}
\address{$^{1}$ GoLP/Instituto de Plasmas e Fus{\~a}o Nuclear - Laborat\'orio Associado, Instituto Superior T\'ecnico, Lisbon, Portugal}
\address{$^{2}$ DCTI, ISCTE - Lisbon University Institute, Portugal}
 \ead{\mailto anne.stockem@ist.utl.pt}

\date{\today}

\begin{abstract}

We assess the impact of non-thermally shock-accelerated particles on the magnetohydrodynamic (MHD) jump conditions of relativistic shocks. The adiabatic constant is calculated directly from first principle particle-in-cell simulation data, enabling a semi-kinetic approach to improve the standard fluid model and allowing for an identification of the key parameters that define the shock structure. We find that the evolving upstream parameters have a stronger impact than the corrections due to non-thermal particles.
%In general, a strong non-thermal tail can change the adiabatic constant significantly and thus yield a measurable modification of the shock properties. However, in the standard case this effect is negligible.
We find that the decrease of the upstream bulk speed yields deviations from the standard MHD model up to 10\%. Furthermore, we obtain a quantitative definition of the shock transition region from our analysis. 
For Weibel-mediated shocks the inclusion of a magnetic field in the MHD conservation equations is addressed for the first time.
\end{abstract}

%\ams{76Y05, 76L05, 85-08, 85A30}

\maketitle

\section{Introduction}

Shocks are common in the universe and a topic of high interest due to their importance in the acceleration of high-energy particles and the subsequent generation of radiation. The most prominent examples are the non-relativistic shocks in supernovae, which can provide an efficient acceleration of cosmic rays inside our galaxy \cite{CB10}, and relativistic shocks in gamma-ray bursts (GRB) \cite{GS09}. 
A clear understanding of the shock properties and their connection to the structure of the fields and the distribution function of the particles is of critical importance to understand and to model many of the scenarios. In particular, as laboratory experiments start to explore in detail these conditions \cite{SM04,tak08,SE10,HT12}
and numerical simulations can capture many of the details of these structures \cite{FH04,CSA08,S08,S08b,MF09,SS09,NN09,H10,fiu111,FF12}, more detailed theoretical models are also required to explain and to predict the properties of relativistic shocks in different contexts \cite{GH92,DB04,SF06}.

The theoretical models to describe the shock properties are based on the hydrodynamic jump conditions, and assume a steady state, neglecting the involved kinetics. In particular, Blandford and McKee \cite{BM76} considered strong shocks, which appear if either the upstream is cold and the energy per particle stays unchanged or if the upstream is ultra-relativistic, so that the rest mass energy can be neglected. In the latter case, energy and pressure are connected by the equation of state \(p = e/3\).
However, and due to the interaction with self-consistent fields in the shock, the particles can be trapped and accelerated in the shock, forming the characteristic high-energy tail in the distribution function, which has been recently reported in simulations (e.\,g.\ \cite{SS09,MF09}). The standard model of the hydrodynamic jump conditions assumes thermal spectra, neglecting the influence of accelerated particles. If the non-thermal tail is strong and the actual particle distribution deviates from such a spectrum, the pressure and energy densities in the downstream vary as well and lead to a modification of the steady state conditions, which can be mathematically expressed by a modification of the adiabatic constant.

In this paper we address the effect of such deviations and derive the jump conditions based on the actual particle distribution in the shock. In particular, we focus on the effects on the shock speed and the density compression ratio which are the key parameters for determining the shock dynamics and energy transport.
We start our analysis with a generalization of the theory for the shock jump conditions for an upstream population with non-zero temperature and discuss the impact of deviations from the idealized contributing parameters on the jump conditions. The theoretical predictions are then compared with fully self-consistent particle-in-cell (PIC) simulations.
Our analysis demonstrates that the modification of the downstream adiabatic constant due to the development of the non-thermal tail as previously reported \cite{CSA08,SS09,MF09} can have a strong impact, but the decrease of the bulk Lorentz factor directly in front of the shock has the dominant influence on the jump conditions. Theory and simulations can be matched for a well-defined shock transition region, thus contributing to identify the different shock regimes.

The analysis has been done for a pure electron-positron plasma, as the expected effects on the adiabatic constant are qualitatively the same as for electron-ion plasmas if the plasma is initially unmagnetized. An initial magnetization suppresses the non-thermal acceleration in pure pair plasmas, and the role of ions becomes then important in this context \cite{S12}.

\section{Theoretical model}
The starting point for the derivation of the shock jump conditions \cite{BM76} are the conservation equations for mass, momentum and energy. We perform our calculations in the downstream rest frame in order to match the configuration of the simulations (see next section). In the standard approach, the one-dimensional strong shock approximation, the upstream is considered to be cold (\(p_1 = 0\)) and the contributions from the self-consistently generated magnetic fields are neglected. Here we include both contributions and follow the formalism of  \cite{zk05}, where quantities with a single index \(Q_{i}\) are measured in their own rest frame and quantities with double indices \(Q_{ij}\) denote the value of species \(i\) in the rest frame \(j\). Throughout the paper, indices 1, 2, s refer to the upstream, downstream, shock frame, respectively. Thus, the conservation equations read
\begin{eqnarray} 
	&& n_1 u_{1s}  = n_2 u_{2s}  \label{Jump_shock1}\\
	&& \beta_{1s} B_{1s} = \beta_{2s} B_{2s}  \label{Jump_shock2}\\
	&& \gamma_{1s} \mu_1 (1+\sigma_1) = \gamma_{2s} \mu_2  (1+\sigma_2)   \label{Jump_shock3} \\
	&& u_{1s} \mu_1 (1\! +\! \frac{\sigma_1}{2\beta_{1s}^2})  \! +\!  \frac{p_1}{n_1 u_{1s} }  = u_{2s} \mu_2 (1\! +\! \frac{\sigma_2}{2\beta_{2s}^2}) \! +\!  \frac{p_2}{n_2 u_{2s} } \,\,\,\,\, \label{Jump_shock4}
\end{eqnarray}
with \(u_{is} = \gamma_{is} \beta_{is}\), where \(\gamma_{is}\) denotes the Lorentz factor, \( \beta_{is} = v_{is} / c\) where  \(v_{is}\) is the bulk velocity, \( \sigma_i = B_{is}^2/(4 \pi n_i \mu_i \gamma_{is}^2) \) is the magnetization, where \(B_{is}\) is the transverse magnetic field, \(n_i\) is the plasma density, and \( \displaystyle \mu_i = 1+ \frac{\Gamma_i -1 }{\Gamma_i} \, \frac{p_i}{n_i} \) is the specific enthalpy. The adiabatic constant \( \Gamma_i\) is defined by the relation between the energy density \(e_i\) and pressure density
%
%\begin{equation}
\mbox{
\(p_i = (\Gamma_i - 1) (e_i - \rho_i)\)
}
%\end{equation}
%
with rest mass density \mbox{\(\rho_i = n_i m c^2\)} .

We start our analysis by considering the case where the magnetic field contribution can be neglected, which is the standard approach for initially unmagnetized shocks \cite{BM76,S08b}. We will later discuss the influence of the self-generated magnetic fields on the jump conditions in the long time evolution of the shock. The shock speed can be determined by performing a Lorentz transformation into the downstream frame and combining equations (\ref{Jump_shock1})-(\ref{Jump_shock4}), yielding
\begin{equation}
	\beta_{s2} = \frac{(\Gamma_2 -1 ) (\gamma_{12} \mu_1 -1)}{\mu_1 \sqrt{\gamma_{12}^2 -1}}
\end{equation}
which depends only on the upstream Lorentz factor \(\gamma_{12}\), the downstream adiabatic constant \(\Gamma_2 \) and the upstream enthalpy \(\mu_1\). A non-zero upstream pressure (\(\mu_1>1\)) increases the shock speed. This effect is weaker the higher the upstream Lorentz factor is and approaches the strong shock approximation for \(\mu_1 = 1\) \cite{S08}.
The density ratio is given by
\begin{equation}\label{betagen}
%\( \displaystyle
	\frac{n_2}{n_{12}} = 1+ \frac{\beta_{12}}{\beta_{s2}} = 1+ \frac{(\gamma_{12}^2-1) \mu_1}{\gamma_{12} (\Gamma_2 -1 ) (\gamma_{12} \mu_1 -1)}
%\)
\end{equation}
and is decreased if the upstream pressure is taken into account. The deviations associated with non-thermal tails will have an impact on the adiabatic constant \(\Gamma_2\).
%%%%%%%%%%% Changes in jump conditions %%%%%%%%%%%%%%%
%
In order to assess the influence of small deviations of the adiabatic constant \(\Gamma_2\) to the typically considered adiabatic constant of an ideal gas \(\Gamma_2^0\), we rewrite the adiabatic constant as \(\Gamma_{2} = \Gamma_2^0 + \delta \Gamma_2\)  where \(\delta \Gamma_2 \ll \Gamma_2^0\). The shock speed is now given by
\begin{equation}
	\beta_{s2} = \beta_{s2}^0 +  \frac{(\gamma_{12} \mu_1 -1)}{\mu_1 \sqrt{\gamma_{12}^2 -1}} \delta \Gamma_2
\end{equation}
and, therefore, the correction of the adiabatic constant increases the shock speed by an amount of the order of \(\delta \Gamma_2\) for a highly relativistic upstream. The density ratio
\begin{equation}\label{eq:density}
	\frac{n_2}{n_{12}}  \approx  \frac{n_2^0}{n_{12}} - \frac{(\gamma_{12}^2-1) \mu_1}{\gamma_{12} ( \Gamma_2^0 - 1)^2 (\gamma_{12} \mu_1 -1 )} \delta \Gamma_2
\end{equation}
is decreased when the correction of the adiabatic constant is included. Typically, an adiabatic constant \( \Gamma_2^0 = 3/2\) for 2D and 4/3 for 3D is used to verify the jump conditions of relativistic shocks, e.\,g.\ \cite{SS09}; therefore, the corrections to the density ratio are of the order of \(4\,\delta \Gamma_2\) in 2D and \(9\,\delta \Gamma_2\) in 3D for a relativistic upstream flow.

Deviations in the Lorentz factor of the flows can also affect the shock jump conditions. Following the previous approach, we define the Lorentz factor of the upstream flow as \(\gamma_{12} = \gamma_{12}^0 - \delta \gamma_{12}\), where \(\gamma_{12}^0\) is the initial Lorentz factor of the upstream flow and \(\delta \gamma_{12}\) its deviation. An increase of \(\delta \gamma_{12}\) reduces the shock speed and enhances the density ratio according to the Taylor expansion of the jump conditions (see \ref{mod:shock:lorentz}).

\begin{figure}[ht!]
\begin{center}
\includegraphics[width=7cm]{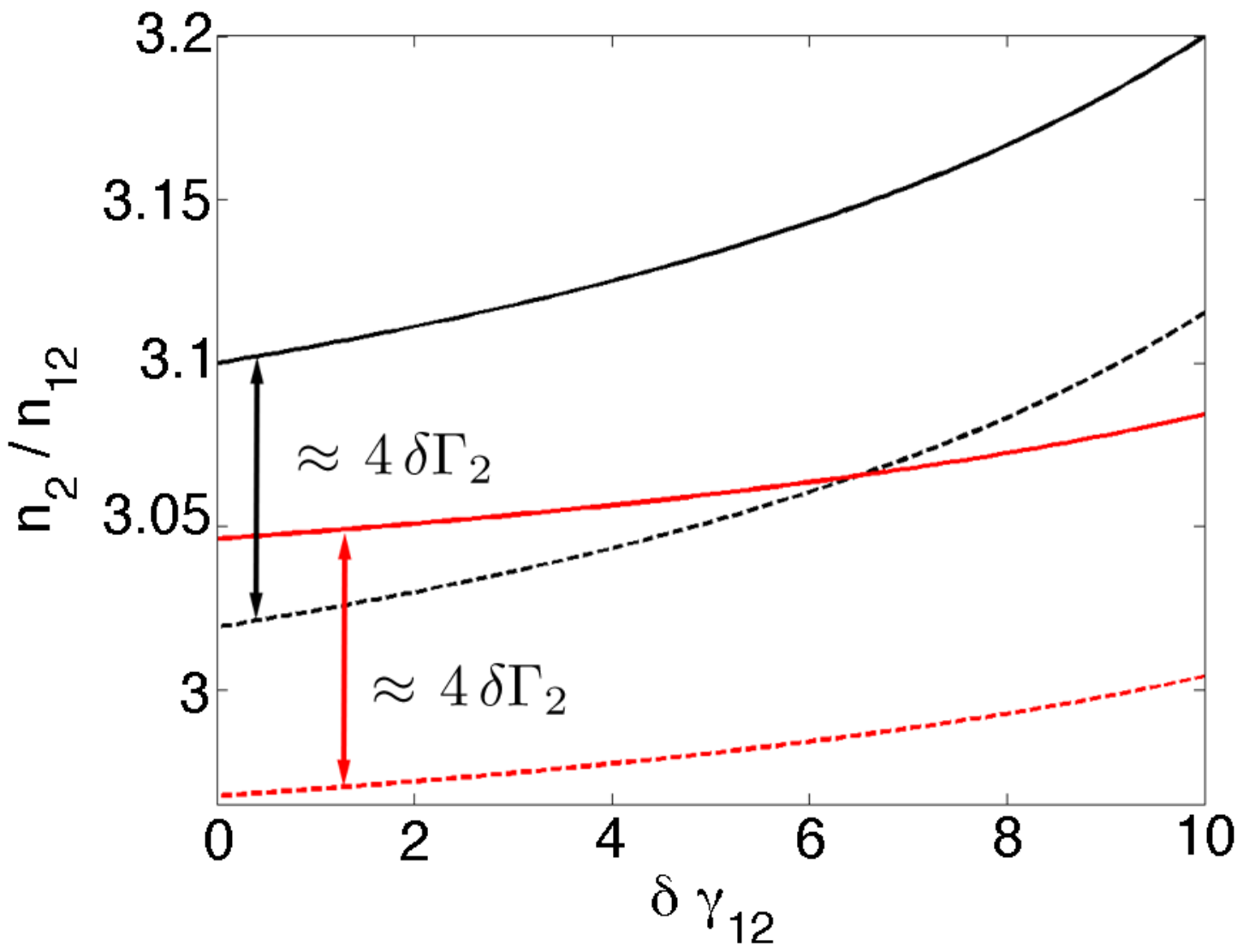}
\end{center}
\vspace{-12pt}
\caption{Effect of the upstream pressure, and of deviations of the adiabatic constant and upstream Lorentz factor in the density ratio. The increase of the upstream pressure and the slowdown of the flow increase the density ratio, whereas deviations on the adiabatic constant decrease the density ratio. All curves are plotted for \(\gamma_{12}^0 = 20\). Black lines correspond to  \(\mu_1 = 1\), red lines to \(\mu_1 = 2\), solid lines to \(\Gamma_2 = 1.5\) and dashed lines to \(\Gamma_2 = 1.52\).}\label{fig:gamma_corrections}
\end{figure}

The effect of the upstream pressure and of deviations of the adiabatic constant and upstream Lorentz factor in the density ratio are illustrated in Figure \ref{fig:gamma_corrections}, summarizing the previous findings and illustrating the stronger impact of the change in the upstream Lorentz factor.

%----------------------------------------------------------------------------
%----------------------------------------------------------------------------

\section{Numerical simulations}
In order to address the effect of the different parameters in realistic scenarios, where the shock structure evolves in time, in a self-consistent manner, we have performed fully relativistic simulations of the shock formation and propagation with OSIRIS 2.0 \cite{F02,F08}. By using a fully kinetic model, the macroscopic quantities describing the shock structure can be calculated directly from the kinetic quantities and compared  with our theoretical model. 
For a given distribution function \(f(\mathbf p)\) obtained from the simulation data, the energy and pressure densities are calculated in the local rest frame of the fluid as
\begin{eqnarray}
	e & := & \frac{\tilde e }{n m c^2} = \int d^3 p \, \gamma \, f(\mathbf p)  \label{energy}\\
	p & := & \frac{\tilde p }{n m c^2} = \int d^3 p \, \frac{p_x^2}{\gamma} \, f(\mathbf p)\label{pressure}
\end{eqnarray}
with \(\gamma = \sqrt{1+ \mathbf p^2}\). Note that the integrals reduce to double integrals for the 2D case. The adiabatic constant can then be calculated from the previously mentioned relation between energy and pressure densities as \( \Gamma = 1+ p/(e-1) \). For a relativistic Maxwellian 
%
%\begin{equation}
\( 
%\displaystyle
	f(\gamma) = C \exp (-\gamma / \Delta \gamma)
\)
%\end{equation}
%
the adiabatic constant yields
%
%\begin{equation}\label{gamma2d}
\mbox{\(
	\Gamma_{2D} = 
	%1+ \frac{p}{e- 1} =
		  (2+3\Delta \gamma)/( 1+ 2 \Delta \gamma )
\)}
%\end{equation}
%
for the 2D case and
%
%\begin{equation}\label{gamma3d}
\mbox{\(
	\Gamma_{3D} =1+ \Delta \gamma/[3 \Delta \gamma -1+ K_1(\Delta \gamma^{-1})/K_2(\Delta \gamma^{-1})]
\)}
%\end{equation}
%
for a 3D geometry with \(K_n(x)\) the modified Bessel functions of the second kind.
The limiting values  are \( \Gamma_{2D} = 2\) for \( \Delta \gamma \rightarrow 0\),  \( \Gamma_{2D} = 3/2\) for \( \Delta \gamma \rightarrow \infty\) and  \( \Gamma_{3D} = 5/3\) for \( \Delta \gamma \rightarrow 0\),  \( \Gamma_{3D} = 4/3\) for \( \Delta \gamma \rightarrow \infty\).

We can immediately observe that, assuming a full thermalization of the upstream flow in the downstream, with a spread \(\Delta \gamma = (\gamma_{12}-1) / 2\) \cite{S08b}, the density ratio equation (\ref{betagen}) reduces to \(n_2/ n_{12} = 3\) in 2D, independent of the initial upstream Lorentz factor. Even for a highly relativistic flow, the deviations arising from the correction of the adiabatic constant can be noticeable, for instance \(n_2/n_{12} = 3.1\) for \(\gamma_{12} = 20\) and \(n_2/n_{12} = 3.13\) for \(\gamma_{12} = 15\) \cite{CSA08}.

In reality, a more complex distribution function of the particles is expected due to the accelerated particle component. Previous results found the best fit for a Maxwellian bulk plus a power-law tail
\begin{eqnarray}\label{plt_dist}
	f(\gamma)& = & \gamma^{-1}  \frac{dn}{d\gamma}  = C_1 \exp \left[- \gamma / \Delta \gamma \right] \nonumber \\
	 + &&\!\! \! \! \! \!  \! \! \! \! \! C_2 \gamma^{-\alpha-1} \min \left\{ 1, \exp \left[-(\gamma- \gamma_{cut} ) / \Delta \gamma_{cut} \right] \right\} 
\end{eqnarray}
with \(C_2 = 0\) for \(\gamma< \gamma_{min}\) \cite{S08b}. The contribution of this modified distribution to the macroscopic shock properties can be now addressed for the first time, using the self-consistent particle distribution from the simulations. The cumbersome analytical expressions for the energy and pressure densities are presented in \ref{cumbEq}. Large effects are expected for a very strong tail, which is given by a small \(\gamma_{min}\) in combination with a small \(\alpha\). In the case of relativistic shocks, these parameters are such that the contribution from the tail is weak.

%\section{Comparison with simulations}

In our simulations, the relativistic shock is created by injecting a charge neutral electron-positron beam with an isotropic thermal spread of \(10^{-3} c\) and bulk Lorentz factor \(\gamma_{12}^0 = 20\) along the negative \(x_1\) direction. The particles are reflected at the opposite wall and interact with the incoming upstream particles, forming a shock. We use 10000 \(\times\) 300 cells with a resolution \(\Delta x_1 = \Delta x_2 = 0.35 c/ \omega_p\), where \( \omega_p = \sqrt{4\pi n_{12}^0 e^2 / m_e}\) is the plasma frequency with \(n_{12}^0\) the upstream electron density measured in the downstream frame at \(t=0\). The number of particles per cell is 3 \(\times\) 3, the time step is \(0.25  \, \omega_p^{-1}\), and the total simulation time is \(4800 \, \omega_p^{-1}\).

%----------------------------------------------------------------------------
\begin{figure}[ht!]
\begin{center}
\includegraphics[width=10cm]{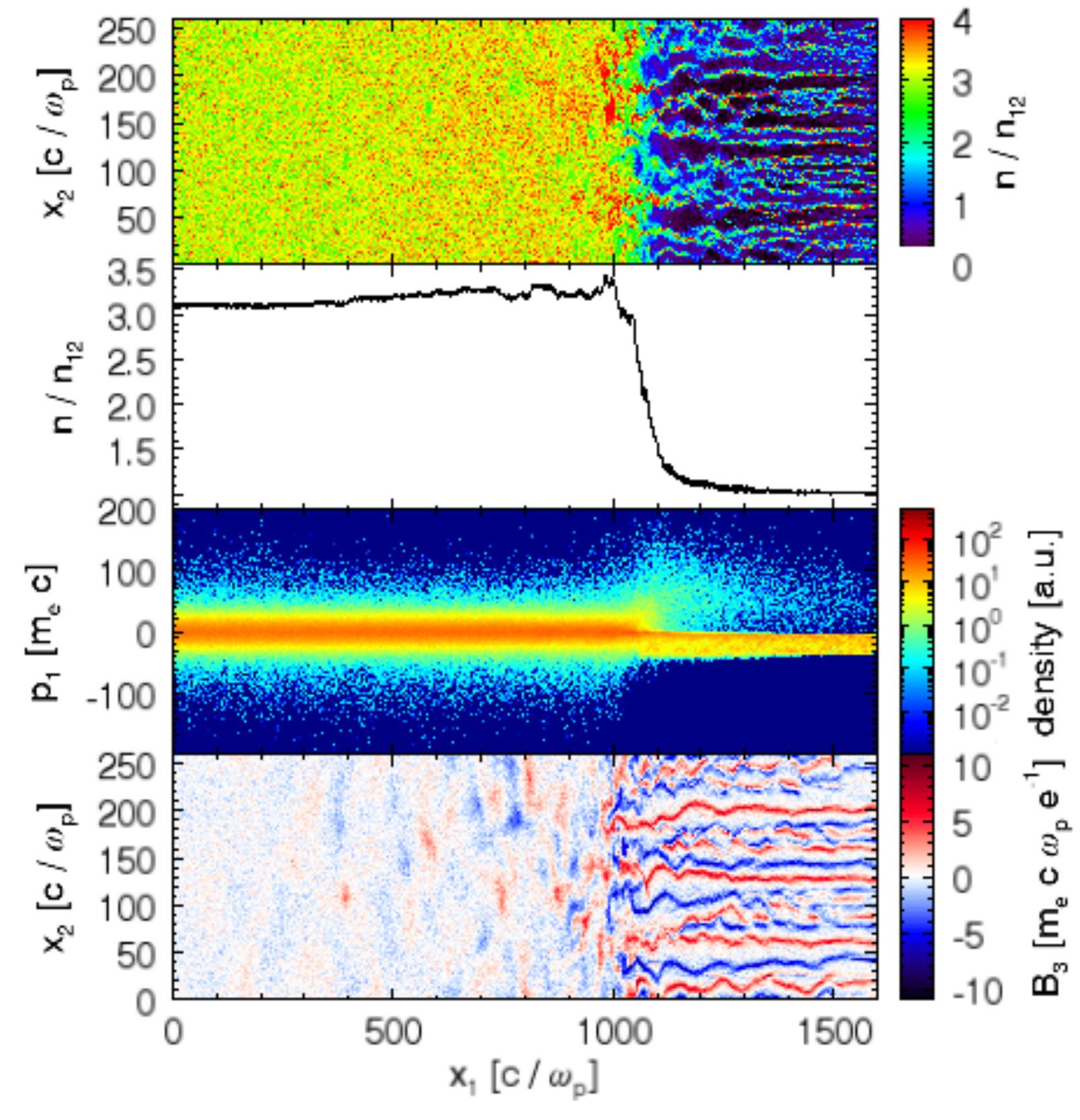}
\end{center}
\vspace{-12pt}
\caption{Shock structure at \( t = 2395 \, \omega_p^{-1} \): the charge density in the \(x_1\)-\(x_2\) plane (a), its spatial average (b), the phase space diagram (c) and the \(B_3\) component of the magnetic field (d).}\label{fig:densities}
\end{figure}
%----------------------------------------------------------------------------

The shock, which propagates along the positive \(x_1\) direction, is formed after \( t \approx 350 \,\omega_p^{-1} \). Figure \ref{fig:densities} shows the important physical quantities at \( t = 2395 \, \omega_p^{-1} \). The typical filamentary structure of Weibel-mediated shocks ahead of the shock front can be seen in the charge density as well as in the magnetic field (figure \ref{fig:densities}a, d) and the density compression factor is \( \approx 3\) (figure \ref{fig:densities}b). The phase space diagram (figure \ref{fig:densities}c) shows the thermalized downstream region on the left hand side and the shock transition region with escaped and reflected particles on the right hand side of the shock front (located around \(1000 \, c/ \omega_p \) for the conditions of Fig.\ \ref{fig:densities}. 

At early times, the filamentary structure does not affect significantly the shock structure and its influence at later times is addressed in the following section. However, averaging over the transverse spatial component gives good qualitative agreement between the theoretical estimates and the simulation results throughout the entire shock propagation.

\section{Discussion}

\begin{figure}[ht!]
\begin{center}
\includegraphics[width=7cm]{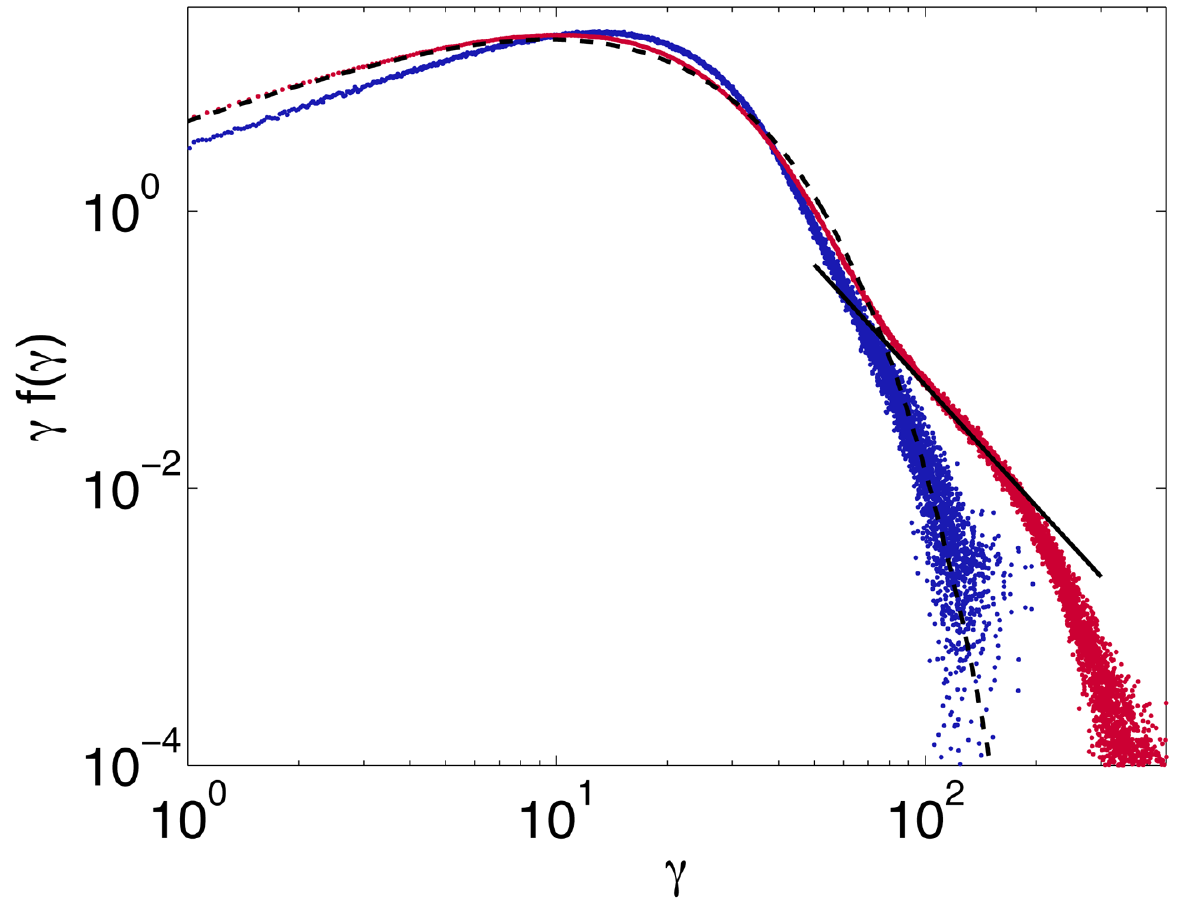}
\includegraphics[width=7cm]{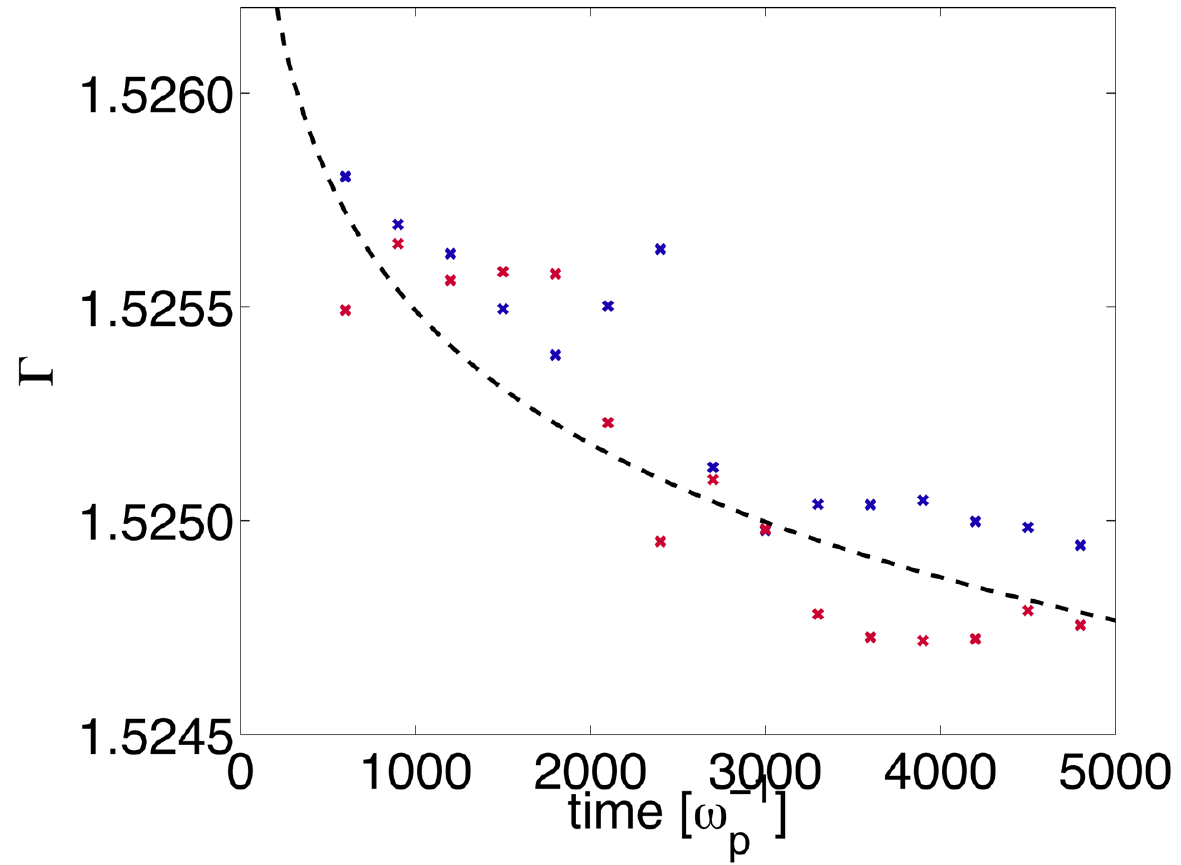}
\end{center}
\vspace{-12pt}
\caption{(a) Electron distributions for simulation times \(t \omega_p = 900\) (blue), 4800 (red) with Maxwellian fit with \(\Delta \gamma = 9.5\) (dashed) and indication of a power-law with index \(\alpha = -2.9\) (solid black). (b) Evolution of the adiabatic index for electrons (blue) and positrons (red) and logarithmically decaying fit.}\label{fig:distributions}
\end{figure}

The analysis of the density ratio associated with the shock front shows that this ratio can reach up to \(n_2 / n_{12}  = 3.2 \pm 0.08 \). This illustrates that when the shock structure is generated self-consistently the shock density ratio can deviate from the theoretical value \(n_2^0 / n_{12}  = 3 \), which is derived from the jump conditions for a cold plasma \cite{BM76} and a Maxwellian distribution in the downstream with a thermal spread \(\Delta \gamma = 9.5\) (leading to \(\Gamma_2 = 1.525\)). We also observe a slight deviation from the shock velocity \( \beta_{s2}^0 = 0.49\) (\( \beta_{measured} = 0.48\)). Since the impact on the density ratio is clearer, we will limit our detailed discussion to this quantity. In order to analyze the impact of the accelerated particles on the jump conditions we have measured the adiabatic constant directly from the kinetic information of the particles in the simulation data as well as analytically from the fittings to the data in figure \ref{fig:distributions}a. For the analytical {\
 estimate} we assume a particle distribution given by equation (\ref{plt_dist}). Both methods provide essentially the same results. The adiabatic constant decreases logarithmically from initially \(\Gamma_2 = 1.5258\) to \(\Gamma_2 = 1.5247\) at the end of the simulation (figure \ref{fig:distributions}b) which predicts a density change according to equation (\ref{eq:density}) of \(\delta n_2 / n_{12} \approx 0.01\) and does not explain by itself the density deviation which we observe in the simulations.
We note that the changes in the adiabatic constant are very small and the fluctuations of the data points are almost on the same level as the total decrease in \(\Gamma\).

%----------------------------------------------------------------------------
%----------------------------------------------------------------------------

The particle distribution function in the downstream region is almost homogeneous along \(x_1\) and varies slowly, whereas the physics in the shock transition region is highly dynamic. In the following, and in order to calculate the pressure and charge densities along the shock propagation direction, the particle distribution ahead of the shock is treated as a single bulk stream, which might not be appropriate for large simulation times, but in the early stages (up to \(t \omega_{p} \approx 2000\)), the fraction of escaped or reflected particles is low compared to the bulk. The pressure density profile along \(x_1\) is used to define the integration range for the quantities ahead of the shock, the Lorentz factor \(\gamma_{12}\) and the upstream enthalpy \(\mu_1\). The peak in the pressure is considered as the transition between upstream and downstream regions and the integration range is varied up to 300\,c\(/ \omega_p\).
%----------------------------------------------------------------------------
\begin{figure}[ht!]
\begin{center}
\includegraphics[width=9cm]{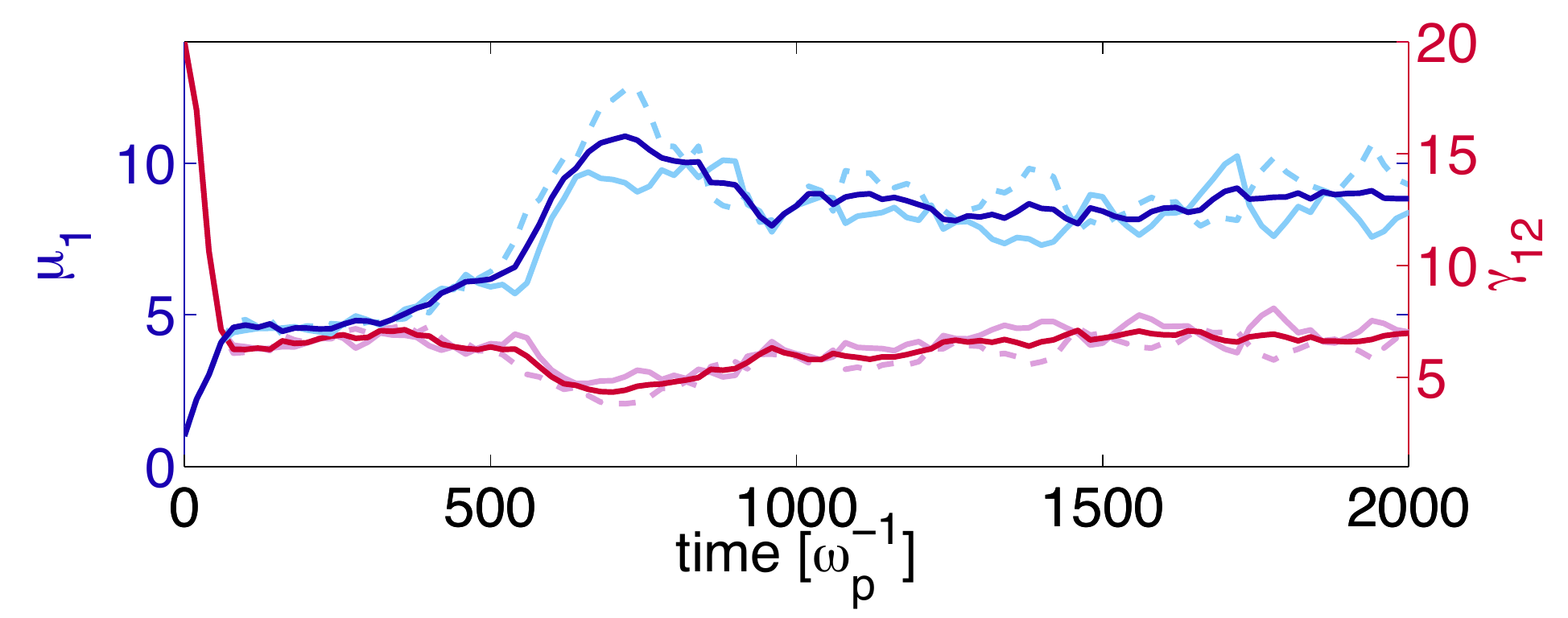}
\includegraphics[width=9cm]{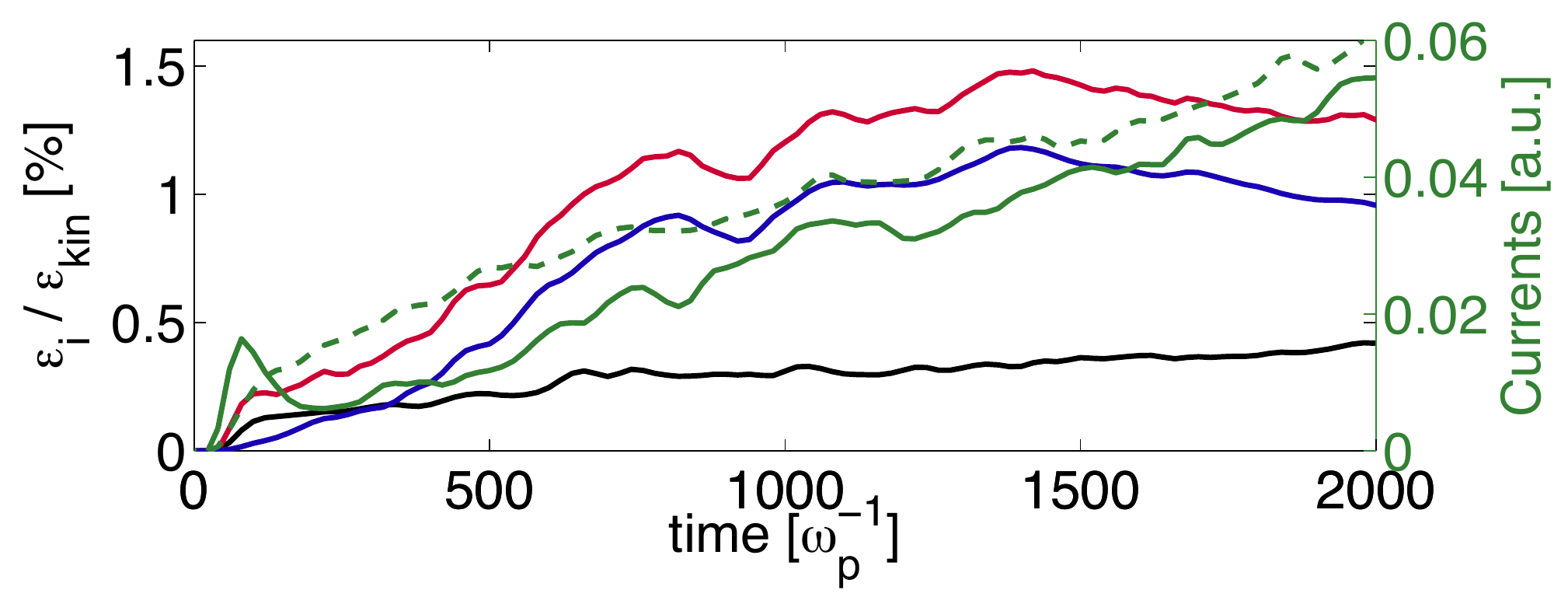}
\end{center}
\vspace{-12pt}
\caption{(a) Specific enthalpy (blue) and average Lorentz factor (red). In light colors, the contribution of positrons (dashed) and electrons (solid) is shown.  The integration region is \(100 \, c / \omega_p\) ahead of the shock front. (b) Normalized field energies \(\epsilon_i / \epsilon_{kin}\) with \(i = B_3\) (red), \(E_1\) (black), \(E_2\) (blue) and currents \(j_1\) (green), \( j_2\) (dashed). \(E_1\) is multiplied by 5.}\label{fig:compar}
\end{figure}
%----------------------------------------------------------------------------

After the shock is formed, the Lorentz factor ahead of the shock deviates strongly from the initial value \(\gamma_{12}^0 = 20\) (Fig. \ref{fig:compar}a), which leads to an increase of the density ratio according to Figure \ref{fig:gamma_corrections}. At the same time, the specific enthalpy has increased, which has a decreasing effect on the density ratio. Both quantities are oscillating in phase, where a high enthalpy appears together with a low bulk Lorentz factor and vice versa. The decrease of the average Lorentz factor in front of the shock stems from a mixing of different populations ahead of the shock: the incoming upstream flow with \(\gamma^0_{12} = 20\), which is decelerated by the fields at the shock front, the particle precursor, which consists mainly of escaping particles that have not been affected by the shock, and the reflected particles from the upstream region. Our simulations reveal that scattering of the flow impinging on the shock front in the self-consistent fields generated in the shock front leads to significant heating (in both the longitudinal direction and in the transverse direction) at the expense of the free energy of the flow, thus contributing to the overall slowdown of the flow as it approaches the shock front.

%----------------------------------------------------------------------------
\begin{figure}[ht!]
\begin{center}
\includegraphics[width=14cm]{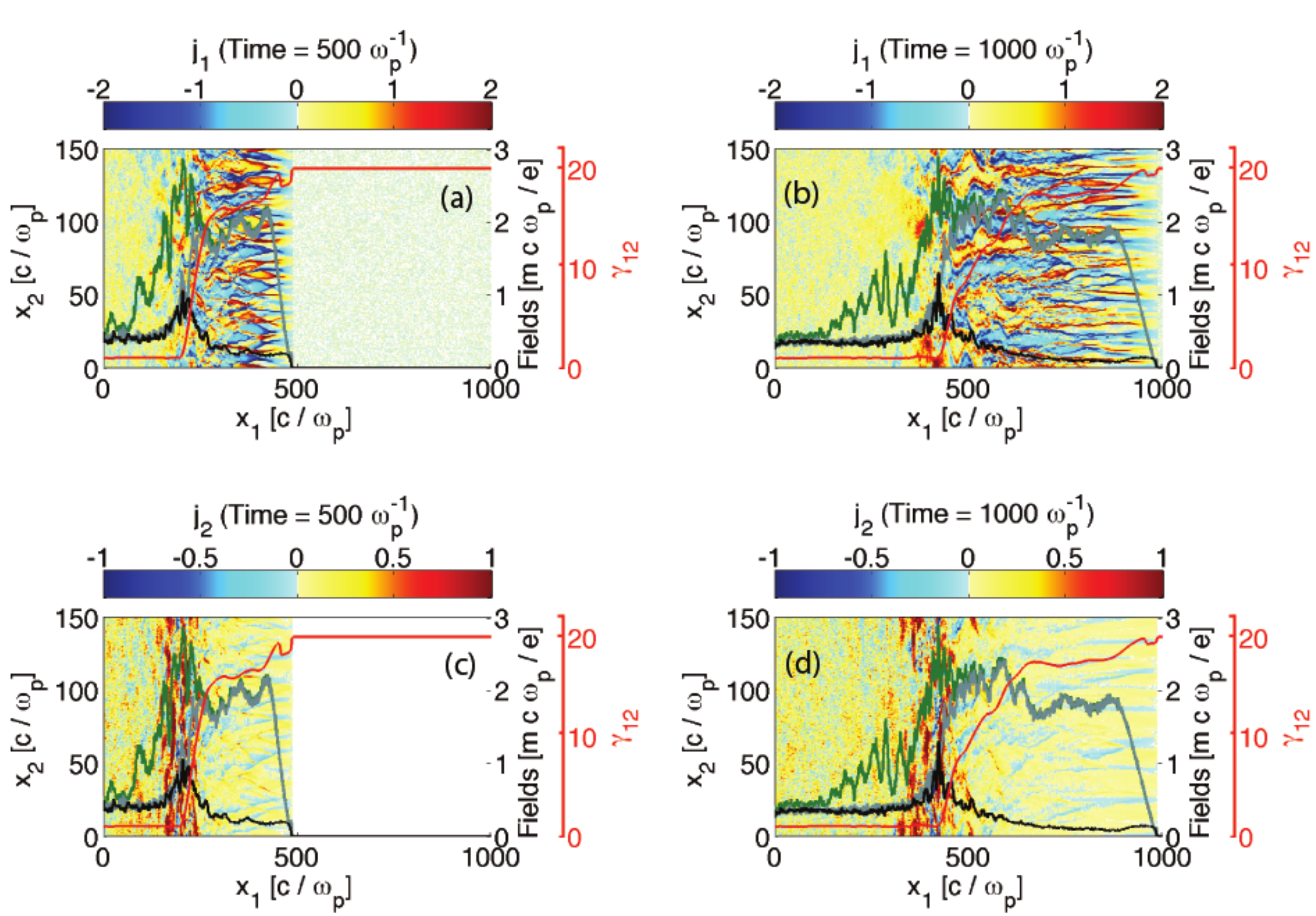}
\end{center}
\vspace{-12pt}
\caption{2D plots: Longitudinal and transverse currents in the shock region at times \(t \omega_p = 500\), 1000. 1D plots: Average fields \(|E_1|\) (black), \(|B_3|\) (green), \(|E_2|\) (gray); \(\gamma_{12}\) (red).}\label{fig:currents}
\end{figure}
%----------------------------------------------------------------------------

Fig. \ref{fig:compar}b shows that the electric field component \(|E_1|\) grows while the Lorentz factor is decreased and reaches a first saturation point at \(\approx 60 \, \omega_p^{-1}\). The growing magnetic field converts energy from the longitudinal momentum \(p_1\) to the transverse component \(p_2\), which causes charge separation between positrons and electrons. The associated current \(j_1\) is responsible for the appearance of \(E_1\), showing its peak value at the same time when the longitudinal field \(|E_1|\) saturates after the linear stage. As the average value \(<\!\!E_1\!\!>\) is zero, statistical changes in the longitudinal electric field component must be responsible for the slowdown of the particles in the shock region.
At \(t \omega_p > 500\) the magnetic field \(B_3\) and the transverse electric field \(E_2\) are increased and the positron and electron species start to oscillate in antiphase around a mean value. \(E_2\) is smaller, but close to \(B_3\) and follows the same trend, which is typical for Weibel-type instability generated filaments.
The transverse fields are generated via instabilities of Weibel-type, which generate and amplify fluctuations in the longitudinal currents. In Fig.\ \ref{fig:currents} the total currents \( j_i = j_{i,e+} + j_{i,e-}\) (\(i = 1,2\)) are plotted at \(t \omega_p = 500\), when the transverse field components start to grow and the oscillations in the species become strong, and at \(t \omega_p = 1000\), when the quantities in Fig.\ \ref{fig:compar}a have reached a quasi-steady state. While \(j_1\) is strong in the entire region of the particle precursor and very weak behind the shock, \(j_2\) exists only in a sharp region around the shock front, and coincides with the peak in \(E_1\).

\begin{figure}[ht!]
\begin{center}
	\includegraphics[width=8.5cm]{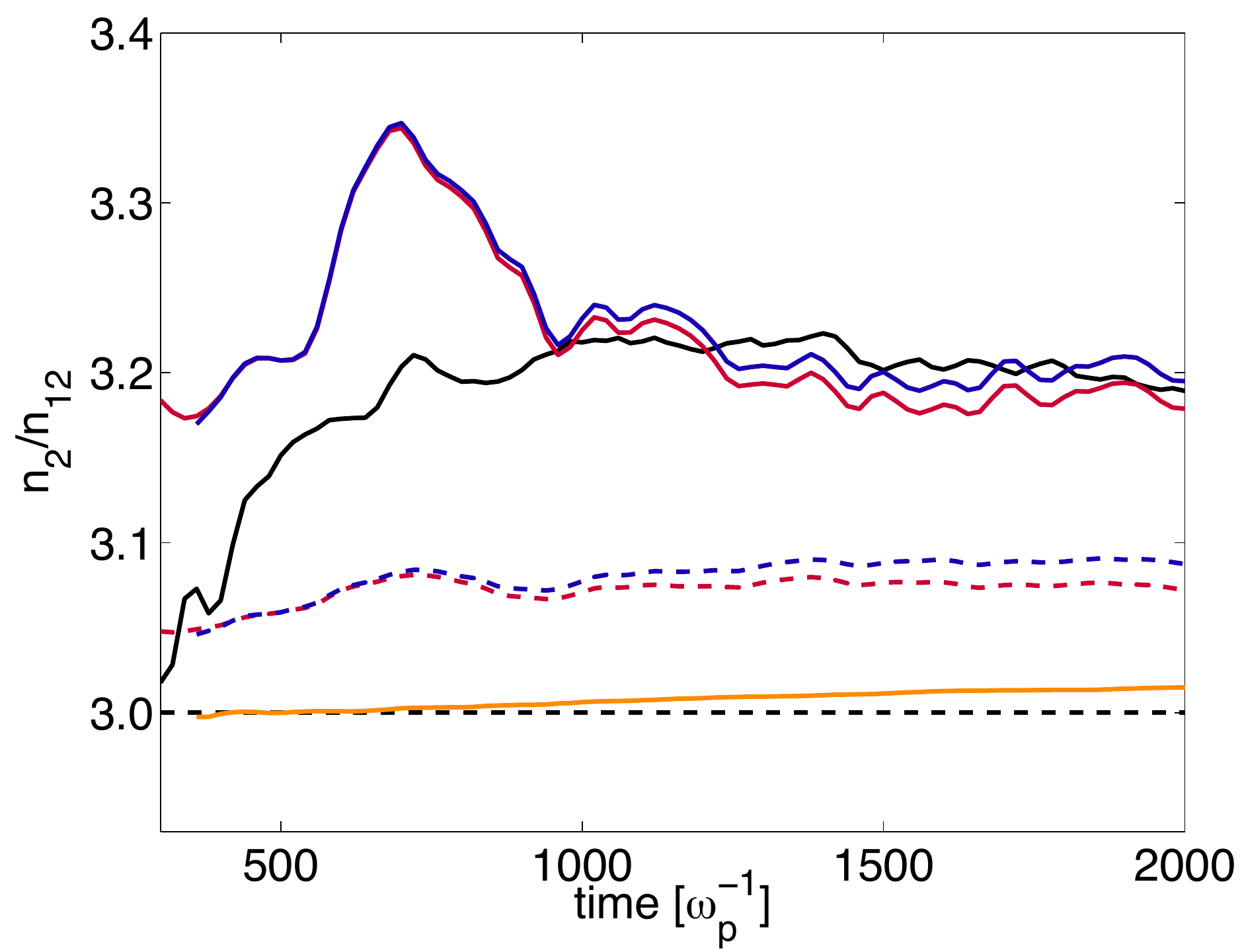}
\end{center}
\vspace{-12pt}
	\caption{Downstream density from simulation data (solid black) and comparison with theoretical models according to table \ref{tab1}: M1 - dashed black, M2 - orange, M3 - dashed red, M4 - dashed blue, M5 - red, M6 - blue.}\label{fig:density}
\end{figure}

Figure \ref{fig:density} compares the average downstream density from the simulation with the different theoretical models listed in table \ref{tab1}. It is clear that the simulation results differ from the ideal model (M1 - no changes in \(\gamma_{12} \) and enthalpy). The inclusion of deviations from a Maxwellian distribution function of the downstream (M2) does not affect the density ratio significantly. On the other hand, the contribution of the decreasing upstream Lorentz factor is observed to have an important impact on the density ratio, but strongly depends on what is defined as the upstream region of the shock. The comparison of the results for different integration ranges shows that after an initial overshoot, the quasi steady state solution of the jump conditions for an integration range of \(100 \, c / \omega_p\) (M5/M6), matches the data best. This suggests that only the vicinity of the shock front within this range significantly affects the shock properties.

%----------------------------------------------------------------------------
%----------------------------------------------------------------------------

%----------------------------------------------------------------------------
%\begin{figure}[ht!]
%\begin{center}
%	\includegraphics[width=8.5cm]{fig6}
%\end{center}
%\vspace{-12pt}
%	\caption{Normalized magnetic field energy integrated over the entire simulation box.}\label{fig:magen}
%\end{figure}
%----------------------------------------------------------------------------

\begin{table}
\begin{center}
\begin{tabular}{c|c|c|c}
	& \(\delta \Gamma_2\) & \(\delta \gamma_{12}\)		&  \(\Delta x_{trans}[c / \omega_p]\)\\
		\hline
		\hline
M1	&  -				& -					& - \\
M2	&  \(\bullet \)	& -					& - \\
M3 	&  - 				&  \(\bullet \)		& 300\\
M4 	&   \(\bullet \) 	&  \(\bullet \) 		& 300 \\
M5 	&  -				&  \(\bullet \) 		& 100\\
M6 	&   \(\bullet \)	&  \(\bullet \) 		& 100 \\
\end{tabular}
\caption{Definition of the models shown in figure \ref{fig:density}: The bullets indicate if the deviations from the adiabatic constant \(\Gamma_2^0= \Gamma_2 - \delta \Gamma_2\) and the Lorentz factor \( \gamma_{12}^0 = \gamma_{12} + \delta \gamma_{12}\) are taken into account and \(\Delta x_{trans}\) denotes the transition region. The bulk is a Maxwellian in all models.}\label{tab1}
\end{center}
\end{table}

If the contributions from the self-generated electromagnetic fields are considered, the resulting density ratio is slightly decreased. In equation (\ref{Jump_shock4}), the first term on the left-hand side and the pressure term on the right-hand side are the dominant terms, and of the same order (\( p_2 / n_2 u_{2s} \approx u_{1s} \mu_1 \approx 20\)). For the magnetization to become important, let us assume a contribution of \(10\%\), so that it has to exceed \(\sigma = 0.05\) as \(\beta_{1s} \approx 1\). The total magnetization in our simulations is \(\approx 0.05\) after a quasi-steady state has been reached, which makes it necessary to be included in the discussion of unmagnetized shocks. The additional decrease of the density ratio due to this contribution is of the order of \(0.1\), which is calculated from the conservation equations (\ref{Jump_shock1})-(\ref{Jump_shock4}).

\section{Conclusions}

In conclusion, we have investigated the evolution of the shock properties when corrections from the usually considered fluid theory are taken into account due to the self-consistent evolution of the shock. We have shown that the shock jump conditions are affected by these corrections, in particular the density ratio.
We found that the formation of a non-thermal tail in the particle distribution and the associated monotonous decrease of the downstream adiabatic constant, as well as the modifications of the upstream bulk speed directly in front of the shock, lead to an increase of the density ratio. The build-up of the upstream pressure and electromagnetic fields have a decreasing effect on the density ratio.

Results from 2D particle-in-cell simulations confirm our theoretical predictions, showing a density ratio 7\% larger than predicted from the standard jump conditions, for early propa\-gation times. The evolution of the upstream Lorentz factor (which has been demonstrated to slow down when approaching the shock \cite{CSA08}) is observed to be the main quantity responsible for such deviations. This analysis allowed us to define the spatial range that determines the shock transition region, which is observed to be 100\,c\(/ \omega_p\), illustrating that the shock is mainly determined by the particles and fields within this range. Our results open the way for a more detailed understanding of the self-consistent evolution of the shock properties, where kinetic effects are taken into account, and demonstrate that a quantitative comparison between shock parameters and simulations/observations should take into account deviations from the standard jump conditions.

\ack
This work was partially supported by the European Research Council (ERC-2010-AdG Grant 267841) and FCT (Portugal) grants SFRH/BPD/65008/2009, SFRH/BD/38952/2007, and PTDC/FIS/111720/2009. Simulations were performed at the IST cluster (Lisbon, Portugal).

\appendix

\section{Changes in the jump conditions due to the Lorentz factor}\label{mod:shock:lorentz}
The full expression of the shock speed is given by 
\begin{equation}
	\beta_{s2} = \frac{(\Gamma_2 -1 ) (\gamma_{12} \mu_1 -1)}{\mu_1 \sqrt{\gamma_{12}^2 -1}}. \nonumber
\end{equation}
To obtain the influence of the upstream Lorentz factor \(\gamma_{12} = \gamma_{12}^0 - \delta \gamma_{12} \), we do a Taylor expansion, yielding
\begin{equation}
	\beta_{s2} = \beta_{s2}^0 -   \frac{(\Gamma_{2} -1) (\gamma_{12}^0 - \mu_1)}{\mu_1 (\gamma_{12}^0 -1)^{3/2}} \delta \gamma_{12}.
\end{equation}
From the density ratio
\begin{equation}
	\frac{n_2}{n_{12}} = 1+ \frac{\beta_{12}}{\beta_{s2}} = 1+ \frac{(\gamma_{12}^2-1) \mu_1}{\gamma_{12} (\Gamma_2 -1 ) (\gamma_{12} \mu_1 -1)} \nonumber
\end{equation}
we obtain
\begin{equation}\label{eq:density}
	\frac{n_2}{n_{12}}  \approx  \frac{n_2^0}{n_{12}} + \frac{ \mu_1 (1 - 2 \mu_1 \gamma_{12}^0 + (\gamma_{12}^0)^2 )}{(\gamma_{12}^0)^2 ( \Gamma_2 - 1) (\gamma_{12}^0 \mu_1 -1 )} \delta \gamma_{12}.
\end{equation}

\section{Analytical expressions for the energy and pressure densities}\label{cumbEq}
For a distribution function consisting of a Maxwellian plus a power-law tail and an exponential cutoff, defined by equation (\ref{plt_dist}), the  analytical expressions for the energy and pressure densities defined in equations (\ref{energy}) and (\ref{pressure}) are given by
\begin{eqnarray}\label{equations1}
	&&e = 2 \pi   \left\{ C_1  \Delta \gamma ( 1+ 2 \Delta \gamma (1+ \Delta \gamma) ) \exp \left( -\Delta \gamma^{-1} \right)   \right. \nonumber\\
	&& \left. + C_2 \left[  \frac{\gamma_{min}^{2-\alpha}- \gamma_{cut}^{2-\alpha}}{\alpha - 2}
	+ \Delta \gamma_{cut}^{2- \alpha} \exp  \left( \frac{\gamma_{cut}}{\Delta \gamma_{cut}} \right)  \Gamma \left( 2- \alpha, \frac{\gamma_{cut}}{\Delta \gamma_{cut}} \right)
	  \right] \right\}\nonumber \\
	&&p = \pi  \left\{2  C_1 \Delta \gamma^2 ( 1+ \Delta \gamma  ) \exp\! \left( -\Delta \gamma^{-1} \right)   %
	 + C_2 \left[ \frac{\gamma_{min}^{2-\alpha} - \gamma_{cut}^{2-\alpha }}{\alpha -2 }  - \frac{\gamma_{min}^{-\alpha} - \gamma_{cut}^{-\alpha }}{\alpha }  \right. \right. \nonumber\\
	 && \left. \left.
	+ \Delta \gamma_{cut}^{-\alpha} \exp\! \left( \frac{\gamma_{cut}}{\Delta \gamma_{cut}}  \right) \! \left[ \Delta \gamma_{cut}^2 \Gamma \!\left( 2- \alpha, \frac{\gamma_{cut}}{\Delta \gamma_{cut}} \right) - \Gamma \!\left( - \alpha, \frac{\gamma_{cut}}{\Delta \gamma_{cut}} \right) \right]  \right] \right\} ,
\end{eqnarray}
where \(\Gamma(n,z)\) stands for the incomplete Gamma function. The constant \(C_2\) is obtained from the simulations, whereas the normalization condition determines
\begin{eqnarray}
	C_1 = \frac{ (2 \pi)^{-1} - C_2 \left[ \exp \left( \frac{\gamma_{cut}}{\Delta \gamma_{cut}} \right) E_\alpha (\Delta \gamma_{cut}^{-1}) - (\alpha -1 )^{-1}\right]}{\Delta \gamma (1+ \Delta \gamma) \exp \left(- \Delta \gamma^{-1} \right) }
\end{eqnarray}
with the exponential integral function \(E_n(z)\).

\bibliography{Bibl_TEX}% Produces the bibliography via BibTeX.

\end{document}